\documentclass[aps,pra,amsmath,twocolumn,superscriptaddress]{revtex4-1}
\usepackage{graphicx}
\usepackage{color}
\usepackage{bm}
\usepackage{amssymb}
\usepackage{xspace}

\newcommand{\fref}[1]{Fig.\ref{#1}}
\newcommand{\neqref}[1]{Eq.\eqref{#1}}
\newcommand{\hc}{$H_{c1}$\xspace}
\newcommand{\hp}{$H_{p}$\xspace}

\newcommand{\tc}{$T_c$\xspace}

\begin{document}

\title{Measurements of the lower critical field of superconductors \\ using NV centers in diamond optical magnetometry}

\author{K.~R.~Joshi}
\affiliation{Ames Laboratory, Ames, IA 50011}
\affiliation{Department of Physics \& Astronomy, Iowa State University, Ames, IA 50011}

\author{N.~M.~Nusran}
\affiliation{Ames Laboratory, Ames, IA 50011}

\author{K.~Cho}
\affiliation{Ames Laboratory, Ames, IA 50011}

\author{M.~A.~Tanatar}
\affiliation{Ames Laboratory, Ames, IA 50011}
\affiliation{Department of Physics \& Astronomy, Iowa State University, Ames, IA 50011}

\author{W.~R.~Meier}
\affiliation{Ames Laboratory, Ames, IA 50011}
\affiliation{Department of Physics \& Astronomy, Iowa State University, Ames, IA 50011}

\author{S.~L.~Bud'ko}
\affiliation{Ames Laboratory, Ames, IA 50011}
\affiliation{Department of Physics \& Astronomy, Iowa State University, Ames, IA 50011}

\author{P.~C.~Canfield}
\affiliation{Ames Laboratory, Ames, IA 50011}
\affiliation{Department of Physics \& Astronomy, Iowa State University, Ames, IA 50011}

\author{R.~Prozorov}
\email[Corresponding author: ]{prozorov@ameslab.gov}
\affiliation{Ames Laboratory, Ames, IA 50011}
\affiliation{Department of Physics \& Astronomy, Iowa State University, Ames, IA 50011}

\date{27 June 2018}

\begin{abstract}
The lower critical magnetic field, $H_{c1}$, of superconductors is measured by using ensembles of NV-centers-in-diamond optical magnetometry. The technique is minimally invasive, and has sub-gauss field sensitivity and sub-$\mu$m spatial resolution, which allow for accurate detection of the vector field at which the vortices start penetrating the sample from the corners. Aided by the revised calculations of the effective demagnetization factors of actual cuboid - shaped samples, $H_{c1}$ and the London penetration depth, $\lambda$, derived from $H_{c1}$ can be obtained. We apply this method to three well-studied superconductors: optimally doped Ba(Fe$_{1-x}$Co$_{x}$)$_2$As$_2$, stoichiometric CaKFe$_4$As$_4$, and high-$T_c$ cuprate YBa$_2$Cu$_3$O$_{7-\delta}$. Our results are well compared with the values of $\lambda$ obtained using other techniques, thus adding another non-destructive and sensitive method to measure these important parameters of superconductors.
\end{abstract}

\maketitle
\section{Introduction}

\subsection{Lower critical magnetic field}

The lower (first) critical field, \hc, is one of the important fundamental parameters characterizing any type-II superconductor \cite{TinkhamBOOK}. Above this field, Abrikosov vortices become energetically favorable and start entering the sample from the edges. Importantly, \hc is related to two fundamental length scales: the London penetration depth, $\lambda$, and the coherence length $\xi$, as follows, \cite{Hu72}

\begin{equation}
H_{c1}= \frac{\phi_0}{4\pi\lambda^2}\left(\ln\frac{\lambda}{\xi}+0.497\right)
\label{Hc1}
\end{equation}
\noindent
$\xi$ enters \neqref{Hc1} only under the logarithm and there are other more direct/sensitive ways to determine it experimentally (for example from the upper critical field, $H_{c2}= \phi_0/\left(2\pi\xi^2\right)$, where $\phi_0=$2.07$\times$10$^{-15}$ Wb is magnetic flux quantum. Thus, the London penetration depth $\lambda$ is often estimated using \neqref{Hc1} if \hc is experimentally given. In terms of the numerical values, for example, for studied here Ba(Fe$_{1-x}$Co$_{x}$)$_2$As$_2$ (122) iron-based superconductors \cite{Canfield10,Prozorov11}, $\xi \approx 2.3$ nm, $\lambda \approx 200$ nm, so that $\kappa = \lambda/\xi \approx 87$, which give $H_{c1} \approx 200$ Oe and $H_{c2} \approx 60$ T. For optimally - doped YBa$_2$Cu$_3$O$_{7-\delta}$ (YBCO) \cite{Gray92,Prozorov00,Sekitani2004}, $\xi \approx 1.6$ nm, $\lambda \approx 140-160$ nm, $\kappa \approx 80-100$, $H_{c1} \approx 350-400$ Oe and $H_{c2} \approx 120$ T.

In practice, using \neqref{Hc1} to determine \hc has two major difficulties: (1) the existence of various surface barriers \cite{Bean64,Brandt95,Brandt99} that inhibit the penetration of a magnetic field, hence lead to over-estimation of \hc, and (2) the distortion of a magnetic field around an actual, finite size sample that leads to under-estimation of \hc. Therefore, the experimentally detected onset of the magnetic field penetration, denoted here \hp, coincides with \hc only in case of an infinite slab in a parallel magnetic field and no surface barrier, a configuration which is almost impossible to achieve in experiment. However, analysis shows that \hp is directly proportional to \hc with the appropriate geometric conversion factor \cite{Brandt99,Brandt01}.  Several previous works analyzed the situation and now most experimentalists follow the numerical results published by E.~H.~Brandt who used approximate nonlinear $E(j)$ characteristics to estimate the connection between \hp and \hc \cite{Brandt99,Brandt01}.

\begin{figure}[tb]
\includegraphics[width=8cm]{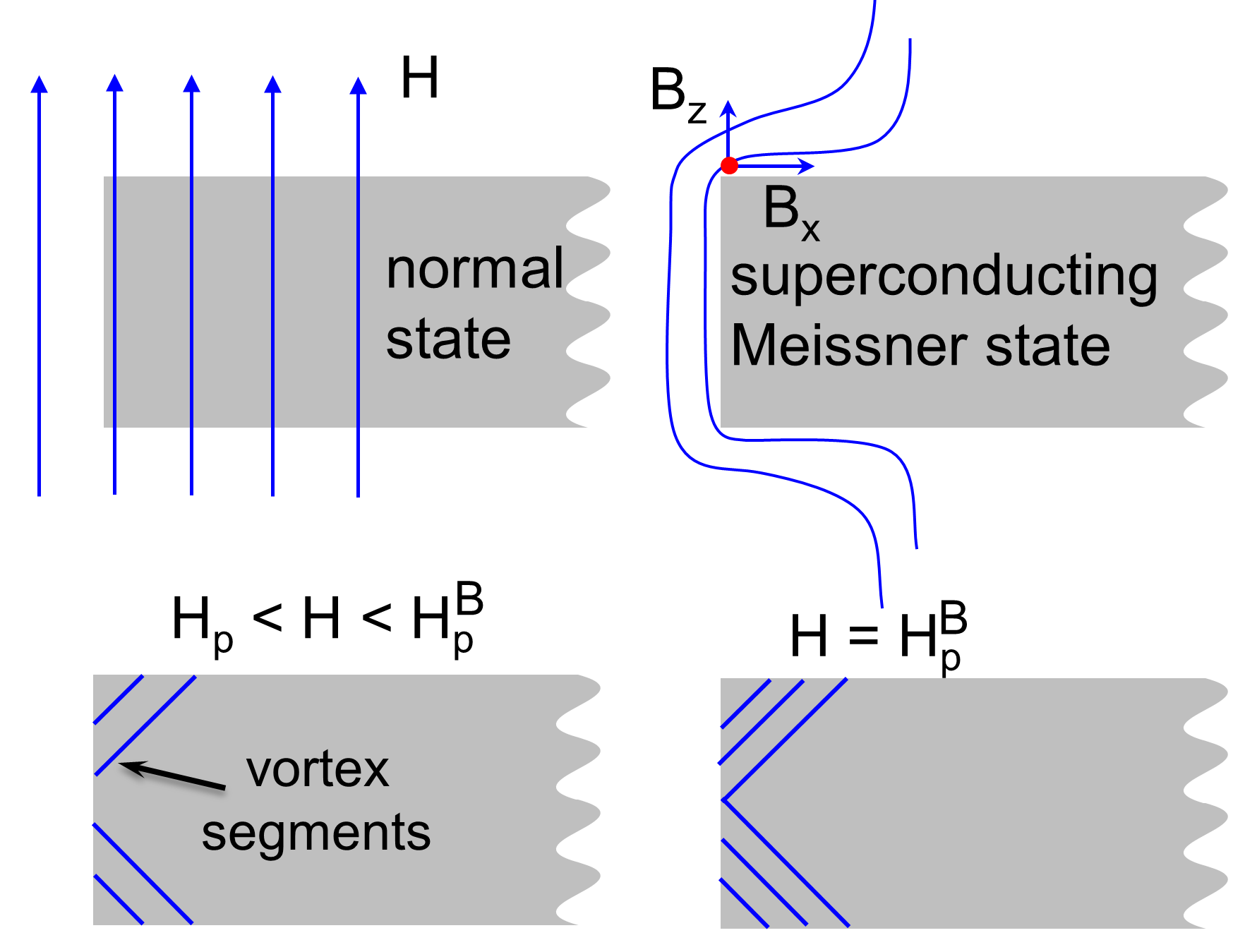}
\caption{Schematics of the magnetic field penetration into a rectangular cross-section sample (only sample side is shown). Top left: normal state. Top right: genuine Meissner state with no vortices and only London penetration depth layer is filled with a magnetic field. Lower left: at $H_p=(1-N)H_{c1}$, Abrikosov vortices start entering the sample from the corners at an angle approximately $45^0$ \cite{Brandt99,Brandt01}. Lower right: at $H=H^B_p$ vortex segments meet in the middle of the edge forming a continous vortex line. Without pinning, these vortices will be immediately moved by the Lorentz force to the sample center.}
\label{fig1}
\end{figure}

Here, it is important to understand how $H_p$ is defined. The E.~H.~Brandt's picture of the magnetic flux penetration into the samples with rectangular cross-sections, known as the ``geometric barrier", is illustrated Fig.~\ref{fig1}. In the normal state (top left panel of Fig.~\ref{fig1}) magnetic field fills nonmagnetic sample uniformly. In the superconducting Meissner state, weak magnetic field is (almost) completely shielded from the interior (top right panel of Fig.~\ref{fig1}), except for the very small layer of London penetration depth, invisible in Fig.~\ref{fig1}. Upon further increase, when maximum local magnetic field $H/(1-N)$, becomes equal to the lower critical field, $H_{c1}$, Abrikosov vortices nucleate and start penetrating the sample. In the present case of a rectangular prism, the maximum field is reached at the corners and straight vortex segments start entering the sample at approximately 45 degrees (bottom left panel of Fig.~\ref{fig1}) \cite{Brandt01}. Since vortex segment length increases, its free energy increases as a function of position from the edge. This is equivalent to a repulsive force that is preventing such vortex from entering the sample. This is the mechanism of a ``geometric barrier" \cite{Brandt99,Brandt01}, which is very different from a Bean-Livingston surface barrier that requires straight smooth surfaces that attract just formed straight vortex lines to their images \cite{Bean64}. When top and bottom segments meet in the middle of the side (at the ``equator", bottom right panel of Fig.~\ref{fig1}) continuous vortex lines are formed. Without pinning, they will immediately be pushed to the sample center by the Lorentz force from the currents due to vortex density gradient. At this value of the applied field, which we denote as $H^B_p$ (``B" emphasizes that this is Brand't definition \cite{Brandt01}), the magnetization, $M(H)$, reaches a maximum amplitude and $H^B_p \approx H_{c1}\tanh{\sqrt{\alpha c/a}}$, where $\alpha=0.36$ for an infinite (in the $b-$direction) strip or $\alpha=0.67$ for disks of radius $a$ \cite{Brandt99,Brandt01}. Note that at this field, $H^B_p$, a significant volume of the sample around its perimeter is already occupied by vortices (from the corner segment penetration) and a local magnetic field at the corners has far exceeded \hc. In fact this the described scenario is the essence of so-called geometric barrier \cite{Brandt01}. (With pinning, above $H^B_p$, vortices will fill the interior according to the critical state model and the $M(H)$ curve will become more non-linear, but will not attain the maximum value until vortices reach sample center.)

An alternative definition of $H_p$ is based on the deviation of local magnetic induction from zero or total magnetic moment from linear $M(H)$ behavior. In practice, the local magnetic induction, $B$, is measured outside the sample, on its surface close to the sample edge. The external magnetic field expelled by the sample leaks into the sensor, so that measured $B(H)$ is always non zero, but is still linear in $H$ and it deviates from linearity when vortices start to penetrate the sample from the corners and this can be detected as the the onset of flux penetration field $H_p$ \cite{Okazaki09,Klein10}. Similar estimate can be obtained from the $M(H)$ curves detecting the deviation from linear behavior upon application of a magnetic field after cooling in zero field \cite{AbdelHafiez13}. Another version of this approach is to look for the remnant flux trapped inside the superconductor which becomes non-zero when a lower critical field is reached in any part of the sample, vortices penetrated and became trapped due to ubiquitous pinnings \cite{Pribulova09}. In all these scenarios, the lower critical field should be  obtained with the appropriate effective demagnetization factor, $N$,
\begin{equation}
H_p=H_{c1}\left(1+N \chi \right)
\label{Hc1N}
\end{equation}
\noindent where $\chi$ is the ``intrinsic" magnetic susceptibility of the material (i.e., in an ``ideal" sample with no demagnetization and surface barriers), which can be taken to be equal to -1 for a robust superconductor at most temperatures below \tc (for an infinite slab of width $2w$ in a parallel field, $\chi=\lambda/w\tanh{\left(w/\lambda \right)}-1$ and it is straightforward to check that $\chi$ is still less than -0.995 even at $T/T_c = 0.99$).

Unfortunately, most previous works that employed local measurements of the onset of magnetic flux penetration using, for example, miniature Hall probes \cite{Klein10,Okazaki09,Pribulova09}, analyzed the data with Brandt's formulas for $H^B_p$ and not with the (more correct in this case) $H_p$ from \neqref{Hc1N}.

\subsection{Effective demagnetizing factors}

To use $H_p$ for determining $H_{c1}$, the effective demagnetizing factor, $N$, has to be calculated for specific sample geometry. Indeed, strictly speaking, $N$ is only defined for ellipsoidal samples, which is of little practical use for typical samples of a cuboidal (rectangular prism) shape. Yet, it is possible to introduce effective demagnetizing factors which were calculated in several previous works, including the cited Brandt's papers, since his estimate of $H^B_p$ implicitly includes the effective $N$ \cite{Brandt01}.
As we recently showed from a full 3D finite-element analysis \cite{Prozorov18}, Brandt provided very accurate expressions for demagnetizing factors in cases of infinite strips or disks of rectangular cross-section, see Eq.(7) in Ref.\cite{Brandt01}. However, we also found that the effective demagnetizing factors for finite cuboids are quite different from the infinite 2D strips and, therefore, the whole methodology of estimating $H_{c1}$ from magnetic measurements should be revisited. This is the subject of the present work.

Although we can calculate the effective demagnetization factor with arbitrary precision for a sample of any shape, it is always useful to have simple, but accurate enough formulas \cite{Prozorov18}. A good approximation for a $2a \times 2b \times 2c$ cuboid in a magnetic field along the $c-$direction is given by \cite{Prozorov18},%
\begin{equation}
N^{-1}=1+\frac{3}{4}\frac{c}{a}\left(1+\frac{a}{b}\right) %
\label{Ncuboid}
\end{equation}

Having samples of rectangular cross-section is problematic from the uncertainty in demagnetization effects point of view, but it is advantageous in terms of the (absence) of surface barriers, because now magnetic flux penetrates from the corners and not parallel to the extended flat surfaces which is how surface barriers are formed \cite{Bean64}. Moreover, the ``geometric barrier" that essentially involves the flux corner penetration described above \cite{Brandt99,Brandt01} is not relevant if the onset of nonlinearity is detected near the sample edge.

\begin{figure}[tb]
\includegraphics[width=9cm]{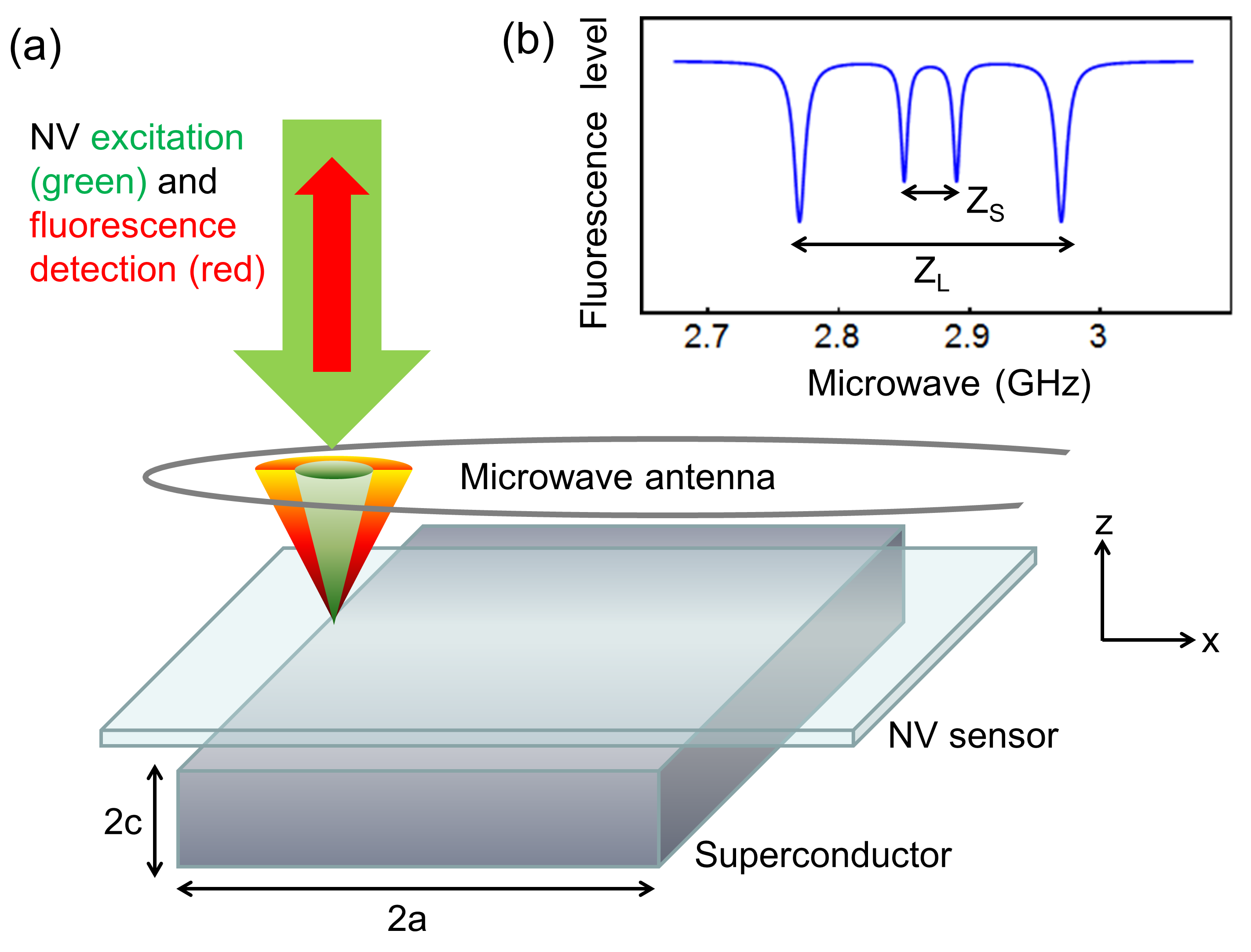}
\caption{(a) Schematics of the key components of the NV sensing setup (b) Optically detected magnetic resonance (ODMR) spectrum for local magnetic field vector with two components, $\vec{B}=(B_x,0,B_z)$. (See text for details.)}
\label{fig2}
\end{figure}

\section{Experimental}
\subsection{Optical magnetic sensing\ using\ \\ NV centers in diamond}

In this work, the vector magnetic induction on the sample surface was measured using optical magnetometry based on nitrogen-vacancy (NV) color centers in diamond. Specifically, the optically detected magnetic resonance (ODMR) of Zeeman split energy levels in NV centers, proportional to a local magnetic field, is measured \cite{Rondin13}. The NV-centers' magneto-sensing has several important advantages for measurements of delicate effects in superconductors. (1) It is minimally invasive, - the magnetic moment of the probe itself is of the order of a few Bohr magnetons, $\mu_B$, and hence has negligible effect on the measured magnetic fields. (2) It has sufficient spatial resolution, - sub-micrometer spatial mapping can be achieved even with the ensemble mode of NV sensing. (3) It is capable to measure a vector magnetic induction \cite{Nusran18}. This is particularly important as the detection of flux penetration depends on the location, and magnetic field lines deviate significantly from the direction of the applied field \cite{Prozorov18}.

Measurement protocols, experimental schematics and deconvolution of the ODMR spectrum into magnetic field components are discussed in detail in our previous work in which the spatial structure of the Meissner state in various superconductors was studied \cite{Nusran18}. Here, we focus particularly on the measurements of the lower critical field, \hc, and summarize the key experimental details for completeness.

To measure a local magnetic induction, a magneto-optical ``indicator" ($1.5 \times 1 \times 0.04$ mm$^3$ diamond plate with embedded NV centers) is placed on top of the superconducting sample with its NV-active side facing the sample surface. On the ``active" side, NV centers are created within $\sim$~20~nm from the surface of a single crystalline diamond plate using commercial protocols that involve nitrogen ion implantation, electron irradiation and high temperature annealing in high vacuum. The diamond plate has $(100)$ crystal surface and $[100]$ edges. Therefore, NV centers are oriented along all four $[111]$ diamond axes, which define the directions of the magnetic field sensing. As a result, possible Zeeman splittings in a random ensemble of NV centers in (indeed, a single crystal of) diamond is given by $2\gamma_e |\vec{B}\cdot\hat{d}|$,  where $\gamma_e \approx 2.8$ MHz/G is the gyromagnetic ratio of the NV-center electronic spin, and $\hat{d}$ is a unit vector along any of the four diamond axes. In a magnetic field along the $\hat{z}$ direction, i.e., $\vec{B}=(0,0,B_z)$, all possible NV orientations result in the same splitting,
\begin{equation*}
Z = \frac{2 \gamma_e B_z}{\sqrt{3}} ~ \approx ~ 3.233\,\ \mathrm{MHz/G}
\end{equation*}
\noindent whereas, if the magnetic field has two components such that $\vec{B}=(B_x,0,B_z)$, the NV ensemble will result in two pairs of Zeeman splittings:
\begin{equation*}
Z_{L,S}  =  Z|B_z \pm B_x|
\end{equation*}
\noindent where, $Z_L$ ($Z_S$) refers to larger (smaller) Zeeman splitting. An example of such two-pairs of ODMR splitting spectrum is shown in \fref{fig2}(b).

\subsection{Experimental setup and samples}
The experimental setup is based on the Attocube CFM/AFM system and includes a confocal microscope optimized for the NV fluorescence detection inside the helium cryostat with optical parts in vacuum and the sample placed on a temperature-controlled cold stage. A schematic of the experiment is shown in \fref{fig2}(a). The objective is focused on the NV centers in a (optically transparent) diamond plate, so that the convolution of the diffraction limited confocal volume with the NV distribution leads to essentially a disk shaped sensing volume of thickness $\approx 20$~nm and diameter $\approx 500$~nm. The diamond plate is placed directly on top of a flat sample surface covering the edge and with NV active side facing the sample. A $50\times$ confocal microscope objective is used both for green laser excitation and red fluorescence collection. Microwave radiation with a very small amplitude is applied using a single-turn 50~$\mu$m diameter silver wire.

All samples were pre-characterized using various thermodynamic and transport techniques (see, e.g., Ref.~\cite{Cho17}) and imaged using scanning electron microscopy (SEM) and only samples with well-defined surfaces and edges, as shown in \fref{fig3}(a), were selected for further measurements.

\section{Measurements of the lower critical field}
\begin{figure}[tb]
\includegraphics[width=9cm]{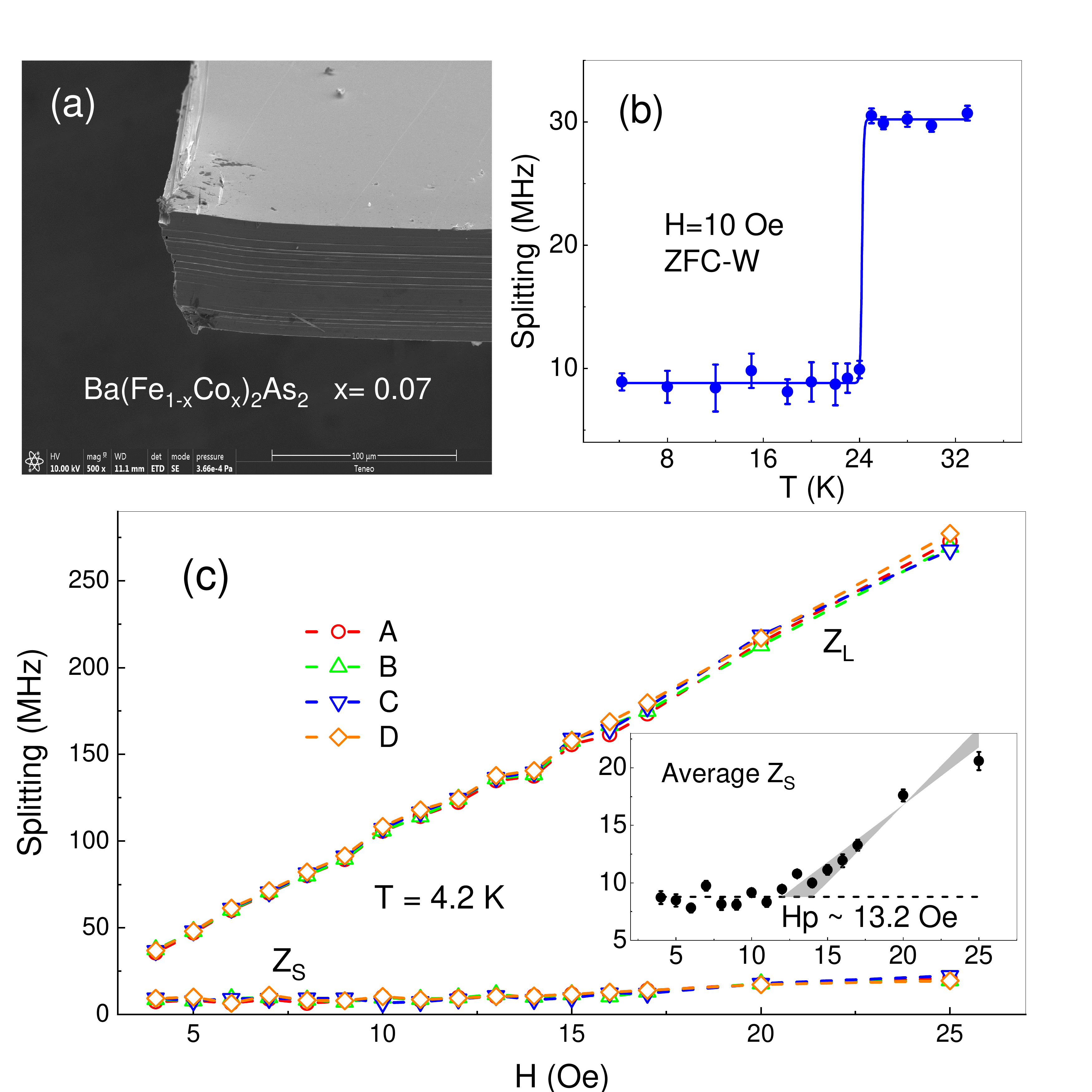}
\caption{(a) Scanning electron microscope (SEM) image of the measured single crystal of Ba(Fe$_{1-x}$Co$_{x}$)$_2$As$_2$, x=0.07 (b) Detection of superconducting phase transition at $T_c\approx$~24~K (c) \hc measurements of this sample at 4.2 K. Zeeman splittings measured at four different points, A, B, C and D near the edge as a function of the increasing magnetic field applied after ZFC. Inset shows a clear change at $H_p$=13.2$\pm$1~Oe of the 4-point-averaged signal of the Z$_S$}
\label{fig3}
\end{figure}

The experimental protocol for measurements of \hc is as follows:

(1) The sample is cooled to the target temperature below \tc in the absence of a magnetic field (zero-field cooling, ZFC). Then, a small magnetic field (10 Oe in our case, much smaller than 200 - 400 Oe expected for \hc at low temperatures as discussed in the introduction) is applied and ODMR signals are recorded at different points along the line perpendicular to the sample edge. Measured ODMR splittings are then converted into the magnetic induction values as described above. This, combined with direct visualization of the sample through a transparent diamond plate, allows for accurate determination of the location of the sample edge and provides information about sample homogeneity. The quality of the superconductor is also verified by looking at the sharpness of the transition detected by the ODMR splitting recorded as a function of temperature at any fixed point over the sample, see, e.g., \fref{fig3}(b).

(2) After this initial preparation and edge identification, the magnetic field is removed, the sample is warmed up to above \tc and then cooled back down to a target temperature, thereby resetting it to the genuine superconducting state with no trapped magnetic field inside. A point inside and over the sample, but close to the edge, is chosen and ODMR spectra are recorded as a function of external magnetic field, which is applied incrementally in small steps. At each step, the superconducting magnet is switched to a persistent mode to insure stability of the magnetic field. The deviation from the linear behavior in $Z_S$ is then detected and recorded as the field of first flux penetration, $H_p$.

(3) Now, using \neqref{Hc1N}, \eqref{Ncuboid}, and \eqref{Hc1}, the value of \hc and the London penetration depth $\lambda$ are evaluated. This procedure is repeated at several locations along the edge to ensure objectivity of the result.

\section{Results and Discussion}
To illustrate the described method, we measured \hc and evaluated the London penetration depth, $\lambda$, in three different superconducting materials.

\subsection{Ba(Fe$_{1-x}$Co$_{x}$)$_2$As$_2$, $x = 0.07$}
A well characterized optimally doped single crystal of Ba(Fe$_{1-x}$Co$_{x}$)$_2$As$_2$, $x = 0.07$ (FeCo122) of cuboidal shape with dimensions, $1.0 \times 1.2 \times 0.05$ mm$^3$, was selected. An SEM image in \fref{fig3}(a) shows a well-defined prismatic corner with flat clean surface and straight edges. The superconducting transition temperature, $T_c \approx 24$~K, determined from a conventional magnetometer, was also consistent with our ODMR measurements at the location on the sample surface inside the sample as shown in \fref{fig3}(b). ODMR splittings at four different locations on the sample surface near the edge are labeled A, B, C, and D in \fref{fig3}(c). These four points are approximately 5 $\mu$m far apart from neighbor points and each point is approximately 10 $\mu$m from the edge inside the sample. As discussed above, the two Zeeman splittings Z$_L$ and Z$_S$ correspond to linear combinations of horizontal (B$_x$) and vertical (B$_z$) components of the magnetic induction as described above. Notice excellent reproducibility of the results indicating homogeneous superconducting properties of our sample. The inset \fref{fig3}(c) shows average (of four points) small splitting signal (Z$_S$). A clear onset of first flux penetration is determined at at $H_p$=13.2$\pm$1~Oe.

To understand the observed ODMR splittings, we consider Brandt's results of flux corner cutting and entering in form of Abrikosov vortices approximately at an angle of $45^0$ with respect to the corner. Therefore the normal to the sample surface $z$-component (along the applied field) and longitudinal, $x$-component of the magnetic induction are approximately equal and proportional to the applied field. This linear relation continues with the increasing applied field until a critical value of the first flux penetration field, $H_p$, is reached. At this point, angle of the magnetic flux at the sample edges deviates from $45^0$ trending more  towards $\hat{z}$ direction. This scenario can be phenomenologically modeled by representing the magnetic induction components as: $B_{z,x}  =  D H \pm \delta$ and $\delta  = 0 + \alpha\theta(H-H_p) (H-H_p)^n$ where $D$ is an effective demagnetization factor and $\theta (H)$ is a Heaviside step function. Because the larger splitting -$Z_L$ and smaller splitting -$Z_S$ are proportional to the sum and difference of $B_{z,x}$ components respectively, the change at $H_p$ is reflected clearly in $Z_S$ but not in $Z_L$. The Zeeman splittings observed in \fref{fig3}(c) can be understood with this model for the parameters: $D=3.5$, $H_p=13.2$, $\alpha=0.6$, and $n=1$. Hence, this provides an experimental confirmation for Brandt's description of flux corner cutting and entering approximately at an angle of $45^0$ with respect to its sides.

\begin{table*}[tb]
\centering
    \begin{tabular}{ccccccc}
    \hline
    \hline
    Superconductor~~~&~~~$T_c$~(K)~~~&~~~$H_{c1}^{2D}$~(G)~~~&~~~$\lambda^B_{ab}$~(nm)~~~&~~~$H_{c1}$~(G)~~~&~~~$\lambda_{ab}$~(nm)~~~&~~~$\lambda$~(nm)~(literature) \\
    \hline
     FeCo122         & 24.3 & 102$\pm$8  & 288$\pm$12 & 158$\pm$12 & 226$\pm$10  & 270,245,224~\cite{Gordon10, Luan11, Williams10}\\
     CaKFe$_4$As$_4$ & 34   & 139$\pm$18 & 251$\pm$18 & 394$\pm$52 & 141$\pm$11  & 208,187~\cite{Khasanov18}\\
     YBCO            & 88.3 & 163$\pm$15   & 236$\pm$12 & 344$\pm$31 & 156$\pm$8 & 146,160,155,149~\cite{Prozorov00,Wu89,Tallon95,Sonier94}\\
    \hline
    \hline
    \end{tabular}
\caption{Estimates for\hc and $\lambda_{ab}$. Here $\lambda^B_{ab}$ is calculated using Brandt's formula for $H^B_p$ in an infinite strip.}
\label{hc1table}
\end{table*}

\begin{figure}[tbh]
\centering
        \scalebox{1}[1]{\includegraphics[width=8cm]{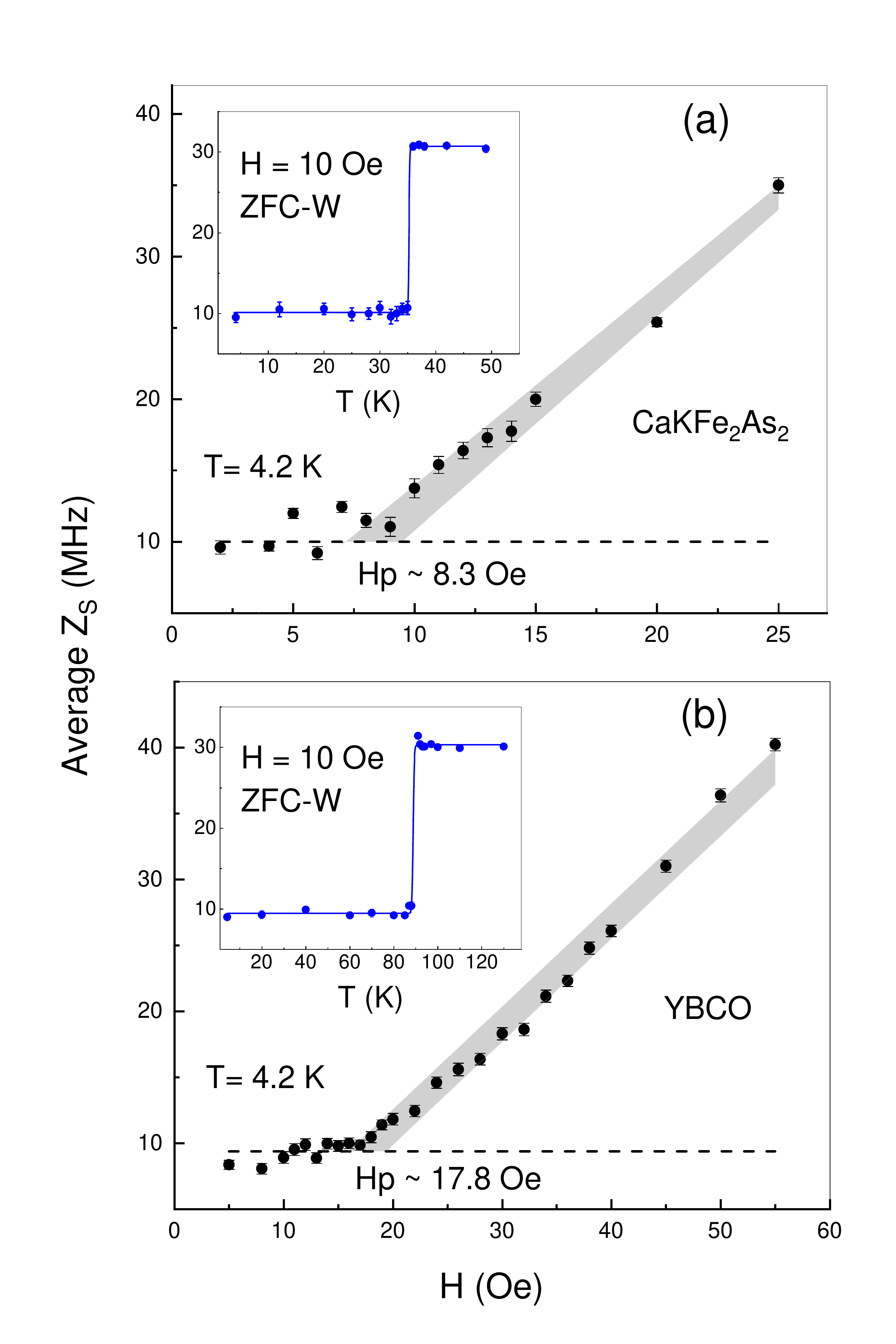}}
        \caption{Measurements of the field of first flux penetration, $H_p$, in single crystals of (a) CaKFe$_4$As$_4$ and (b) YBa$_2$Cu$_3$O$_{7-\delta}$. Insets show superconducting phase transitions at $T_c \approx 34$ K and $88$ K, respectively.}
        \label{fig4}
\end{figure}

From the experimental value of $H_p$ and effective demagnetization factor for this particular sample, $N=0.9168$ we obtain using \neqref{Hc1N}, \hc=158$\pm$12~Oe. And with the use of \neqref{Hc1} and taking $\xi \approx 2.3$ nm, we obtain the final result, $\lambda = 226\pm$10~nm. This estimate for London penetration depth is comparable with the values obtained from other techniques such as $\mu$SR - 224~nm \cite{Williams10} and MFM - 245~nm \cite{Luan11}. The agreement is quite good and gives confidence in the validity of the developed technique. Table~(\ref{hc1table}) summarize all these estimates. The estimates obtained using Brandt's formula for an infinite rectangular strip are also given for comparison.

\subsection{CaKFe$_4$As$_4$}
The cuboid - shaped single crystal of stoichiometric CaKFe$_4$As$_4$ with dimensions of $1.01 \times 0.99 \times 0.01$ mm$^3$ was studied. The inset in \fref{fig4}(a) shows a sharp superconducting phase transition at $T_c\approx 34$ K. The average of ODMR splitting, Z$_S$,  near the sample edge as a function of the applied magnetic field clearly shows a break associated with the magnetic flux penetration at $H_p = 8.3 \pm 1.1$ Oe. The error here is determined visually by the shaded region which spans all measurement points. Using \neqref{Hc1N} and \eqref{Ncuboid}, this results in the estimation of $H_{c1} = 394 \pm 52$ Oe. Now, using \neqref{Hc1} and $\xi \approx 2.15$ nm \cite{Cho17}, we estimate $\lambda = 141 \pm 11$ nm. This result was used to calculate the superfluid density in Ref.\cite{Cho17}, which was consistent with isotropic two-gap $s_{\pm}$ pairing state.

\subsection{YBa$_2$Cu$_3$O$_{7-\delta}$}
To look at a very different system, we also measured a single crystal of a well known cuprate superconductor, YBa$_2$Cu$_3$O$_{7-\delta}$ (YBCO). The sample dimensions were $0.5 \times 0.85 \times 0.017$ mm$^3$. The inset in \fref{fig4}(b) shows a sharp superconducting phase transition at $T_c \approx 88$ K. The clear break associated with the magnetic field of first flux penetration in the average Z$_S$ vs $H$ plot is observed at $H_p$=17.8$\pm$1.6~Oe. Using \neqref{Hc1N} and \eqref{Ncuboid}, this leads to estimation of $H_{c1} = 344 \pm 31$ Oe. Using \neqref{Hc1} and coherence length $\xi\approx 1.6$ nm \cite{Gray92,Sekitani2004}, we estimate $\lambda \approx 156 \pm 8$ nm. All estimates including values obtained using Brandt's formula for a rectangular strip and from other techniques are summarized in Table~(\ref{hc1table}). Once again a good agreement is seen between our estimates and the values reported in the literature obtained from other techniques such as $\mu$SR - 155 nm \cite{Tallon95}, microwave cavity perturbation technique, 160 nm \cite{Wu89} and tunnel-diode resonator, 140 nm \cite{Prozorov00}.

\section{Conclusions}
To summarize, we used NV-centers in diamond for optical vector magnetic field sensing at low temperatures to measure the lower critical field, \hc, in type II superconductors. The minimally-invasive nature and optical diffraction-limited small size of the probe makes NV sensor ideal for this purpose. The capability of resolving vector components provides a unique advantage, which allowed direct verification of the E. H. Brandt's model of magnetic flux penetration that proceeds via corner cutting by vortices at a $\approx 45^0$ angle with respect to the edges.
We applied this technique to three different superconductors: optimally doped FeCo122, stoichiometric CaKFe$_4$As$_4$, and high-$T_c$ cuprate,  YBCO. London penetration depth values evaluated from the obtained \hc are in a good agreement with the literature with the largest uncertainty for CaK1144, most likely due to various levels of scattering in samples studied in different works. Our approach is very useful non-destructive way to estimate $\lambda(0)$ that is needed to obtain superfluid density, which is the quantity that can be compared with theory.

\section*{Acknowledgement}
This work was supported by the U.S. Department of Energy (DOE), Office of Science, Basic Energy Sciences, Materials Science and Engineering Division. The research was performed at Ames Laboratory, which is operated for the U.S. DOE by Iowa State University under contract \# DE-AC02-07CH11358. W. M. was supported by the Gordon and Betty Moore Foundation's EPiQS Initiative through Grant GBMF4411.


%

\end{document}